\documentclass[a4paper,11pt]{article}
\usepackage{graphicx}
\usepackage{mathrsfs}
\usepackage{amssymb}
\usepackage{bm}
\usepackage{latexsym}

\begin{document}
\title{\Large{A classical wave-packet approach to shot noise:\\ power spectrum, Fano factor, and effective charge}}

\author{
        \mbox{}\\
        Hiroyuki Inoue\\
        \small{Braun Center for Submicron Research, Department of Condensed Matter Physics,}\\
        \small{Weizmann Institute of Science, Rehovot 76100, Israel}\\
        \normalsize
            \texttt{hiroyuki.inoue@weizmann.ac.il}
}
\date{}
\maketitle

\begin{abstract}
\noindent The progress in nanofabrication, measurement technology, and mesoscopic transport theory has been expanding the field of shot noise. Although a wave-packet approach to DC shot noise of independent electrons at finite temperature was offered as an intuitive alternative to the sophisticated theories, actual shot noise data often behave more complicated than the derived simple expression. For example, so-called effective charge can deviate from elementary electronic charge due to correlated tunnelings. Also, there are cases where one wishes to know the full spectrum of the shot noise. It will be of great use if a handy method for the shot noise in various experimental situations is available. In this article, a classical wave-packet approach to the shot noise is presented. The classical formulation provides a rigorous yet straightforward formalism to compute the full spectrum and, furthermore, clarifies the structure of Fano factor and effective charge. Additionally, the role of realistic detectors and an application to cross correlation measurements are also discussed. The present method can serve as an intuitive complement to the full quantum mechanical and field theoretical approaches.
\end{abstract}

\newpage
\tableofcontents
\newpage

\section{Introduction}
\subsection{Power spectrum of signal}
Suppose a signal $I(t)$ is of our interest. When the signal spans a long time, the total energy of the signal, $\int_{-\infty}^{\infty}|I(t)|^{2}dt$, may diverge. Hence, the power of the signal is more senseful. Imagine taking a segment of the signal $I(t)$ and name it as $I_{\mathscr{T}}(t)$.
\begin{eqnarray}
I_{\mathscr{T}}(t) & = & I(t) \quad (|t| \leq \mathscr{T}/2) \quad ,\quad 0\quad (|t| > \mathscr{T}/2)
\end{eqnarray}
Here, let us imagine $\mathscr{T}$ being a long enough time compared to any time scale of the system. Its Fourier transform is $\tilde{I}_{\mathscr{T}}(f)$. Since $I(t)$ is a real signal, $\tilde{I}^{*}_{\mathscr{T}}(f) = \tilde{I}_{\mathscr{T}}(-f)$. We will often use $\omega = 2\pi f$ as well. We now consider the auto correlation function of $I(t)$, $C(t)=\langle I(t')I(t'+t) \rangle$.
\begin{eqnarray}
\langle I(t')I(t'+t) \rangle & = & \lim_{\mathscr{T} \rightarrow \infty} \frac{1}{\mathscr{T}}\int_{-\mathscr{T}/2}^{\mathscr{T}/2} dt' I_{\mathscr{T}}(t')I_{\mathscr{T}}(t'+t)\nonumber\\
& = & \lim_{\mathscr{T} \rightarrow \infty} \frac{1}{\mathscr{T}}\int_{-\mathscr{T}/2}^{\mathscr{T}/2} \int\int dt'\frac{d\omega}{2\pi} \frac{d\omega '}{2\pi} \tilde{I}_{\mathscr{T}}(\omega)e^{-i\omega t'}\tilde{I}_{\mathscr{T}}(\omega')e^{-i\omega' (t'+t)}\nonumber\\
& = & \int\int \frac{d\omega}{2\pi} \frac{d\omega '}{2\pi}  \lim_{\mathscr{T} \rightarrow \infty} \frac{1}{\mathscr{T}} \int_{-\mathscr{T}/2}^{\mathscr{T}/2} dt' e^{i(\omega'+\omega) t'} \tilde{I}_{\mathscr{T}}(\omega)\tilde{I}_{\mathscr{T}}(\omega')e^{i\omega' t}\nonumber\\
& = & \int\int \frac{d\omega}{2\pi} \frac{d\omega '}{2\pi}  \lim_{\mathscr{T} \rightarrow \infty} \frac{1}{\mathscr{T}} \Bigg[\frac{e^{i(\omega'+\omega)t'}}{i(\omega+\omega ')}\Bigg]_{\mathscr{T}/2}^{\mathscr{T}/2}\tilde{I}_{\mathscr{T}}(\omega)\tilde{I}_{\mathscr{T}}(\omega ')e^{i\omega ' t}\nonumber\\
& = & \int\int \frac{d\omega}{2\pi} \frac{d\omega '}{2\pi}  \lim_{\mathscr{T} \rightarrow \infty} \frac{1}{\mathscr{T}} \Bigg[\frac{sin(\omega +\omega')\frac{\mathscr{T}}{2}}{\frac{\omega+\omega '}{2}}\Bigg]\tilde{I}_{\mathscr{T}}(\omega)\tilde{I}_{\mathscr{T}}(\omega ')e^{i\omega ' t}\nonumber\\
& = & \int\int df df'  \lim_{\mathscr{T} \rightarrow \infty} \frac{1}{\mathscr{T}} \Bigg[\mathscr{T}\frac{ sin(\pi(f+f')\mathscr{T})}{\pi(f+f')\mathscr{T}}\Bigg]\tilde{I}_{\mathscr{T}}(f)\tilde{I}_{\mathscr{T}}(f')e^{2\pi if ' t}\nonumber\\
& = & \int\int df df' \lim_{\mathscr{T} \rightarrow \infty} \frac{1}{\mathscr{T}} \Big[\mathscr{T} sinc(\pi(f+f')\mathscr{T})\Big]\tilde{I}_{\mathscr{T}}(f)\tilde{I}_{\mathscr{T}}(f')e^{i\omega' t}\nonumber\\
& = & \int\int df df' \lim_{\mathscr{T} \rightarrow \infty} \frac{1}{\mathscr{T}} \delta(f+f')\tilde{I}_{\mathscr{T}}(f)\tilde{I}_{\mathscr{T}}(f')e^{2\pi if' t}\nonumber\\
& = & \int df \lim_{\mathscr{T} \rightarrow \infty} \frac{1}{\mathscr{T}} \tilde{I}_{\mathscr{T}}(f)\tilde{I}_{\mathscr{T}}(-f)e^{2\pi i f t}\nonumber\\
& = & \int df \lim_{\mathscr{T} \rightarrow \infty} \frac{|\tilde{I}_{\mathscr{T}}(f)|^{2}}{\mathscr{T}} e^{2\pi if t}
\end{eqnarray}
$\lim_{\mathscr{T} \rightarrow \infty} \frac{|\tilde{I}_{\mathscr{T}}(f)|^{2}}{\mathscr{T}}$ is called the power spectral density $\bar{S}(f)$. The auto correlation function and the power spectral density are directly connected by the Fourier transform. This relation is called Wiener-Kintchin theorem. Since $I(t)$ is a real function, both $\bar{S}(f) = \bar{S}^{*}(-f)$ contribute to the spectral density at the frequency $f$. Hence, it is natural to redefine the shot noise power spectrum as $S(f)\equiv\bar{S}(f)+\bar{S}(-f)$.
\begin{eqnarray}
S(f) & = & \lim_{\mathscr{T} \rightarrow \infty} \frac{2|\tilde{I}_{\mathscr{T}}(f)|^{2}}{\mathscr{T}}
\end{eqnarray}

\subsection{Definition of shot noise}
The definition of the shot noise is the power spectrum of the current fluctuation, $\delta I(t) = I(t)-\langle I \rangle$, appearing in the presence of driven currents. This is equivalent to the Fourier transform of the second order correlator of current fluctuation. The unit of the shot noise is $A^{2}/Hz$.
\begin{eqnarray}
C(t) & = & \langle \delta I(t') \delta I(t'+t) \rangle\\
S(f) & = & 2\mathscr{F}[C(t)] \quad (f>0)
\end{eqnarray}
In reality, the noise consists of several components such as external, instrumentation, thermal, and shot noises. Especially, the shot noise is the component which appears in the presence of driven current. In the present article, we neglect the external, instrumentation noise. Therefore, the current fluctuation is nothing but the shot noise. Even at finite temperatures, as long as one focuses on the DC voltage bias range beyond the thermal energy, the formulation is directly applicable. Note that $\delta I(t)$ is present even without external bias at an equilibrium finite temperature. This is the origin of the thermal noise, $4k_{B}\Theta G$, where $\Theta$ is the temperature and $G$ is the conductance of the system, which is an example of the fluctuation-dissipation theorem. Here, the fluctuation is $\delta I(t)$ and the dissipation is $G^{-1}$. The thermal noise is known to exhibit a white power spectrum with a high-frequency cutoff at the order of $\frac{2k_{B}\Theta}{h}$

\subsection{What this method is about}
Martin and Landauer employed a wave-packet approach to calculate DC shot noise of finite temperature one-dimensional electrons impinged on a tunnel barrier \cite{Landauer1989}\cite{Landauer1991}\cite{Landauer1992}. Roughly speaking, the shot noise as a function of DC bias exhibits a V-shape with the slope proportional to the charge but rounded only in the small bias range comparable to the temperature. From this, one can extract the temperature and the charge of tunneling entity. Despite of its simplicity, it has been effective to explain various shot noise data even of highly correlated systems such as edge modes of fractional quantum Hall effects \cite{Picciotto1997}\cite{Saminadayar1997}\cite{Moty2006}. Certainly, there are other sophisticated formalisms beyond Landauer's approach. Many experimentalists who just dived into the subject of electronic shot noise may find themselves a bit lost in front of literatures employing a heavy machinery such as Non-equilibrium Green's functions \cite{Haug1996}\cite{Kane1994}. Although it is a powerful method, computing the shot noise in various experimental conditions is often beyond experimentalists' tools (or impossible even for theorists). Furthermore, actual shot noise data can behave in more complicated manners than the simple V-shape. Sometimes one also wishes to know about the full noise spectrum. Hence, it is useful to have a handy method for the shot noise.

This article provides probably the most elementary way to calculate the shot noise in various situations by employing a classical wave packet approach to the shot noise. The mathematical derivations is in a similar form as in the theory of signal processing \cite{BRG1988}. The full expression of the shot noise contains two important terms, the power spectrum of the wave packet and Fano factor. The concept of effective charge will be also discussed. Despite of the simple formulation, it clarifies the structure of the power spectrum, Fano factor and effective charge in a trasparent manner. Therefore, this can serve as a complemental method to the full quantum mechanical or field theoretical formalism.

For clarity, we imagine a one-dimensional conductor and a splitter (partitioner/tunnel barrier) in the middle (e.g. quantum point contacts, quantum dots, etc). Current in such a system can be viewed as a train of wave packets (Fig. 1(a)). In an ordered train of wave packets, neighboring wave packets are separated by a time $\tau$. Each slot is either (multiply) occupied or empty. Important to note that the splitter does not fracture them. Each wave packet maintains its shape in the course of the propagation. Importantly, this method only deals with occupation of each slot and not the probability amplitude of the wave packets. As shown in the Fig. 1(b), the experiment consists of three step: (1) an ordered train of wave packets is driven from a source to the system; (2) the wave packets are partitioned in stochastic or correlated manners into two ways (e.g. transmitted and reflected, or right and left, depending on the realization); (3) each is deteced at detectors 1 and 2 at the end of the system. Here, we start with an ideal detector, which has an infinite bandwidth and is able to resolve the shape of each wave packet and to check the occupation of each time slot. The case of realistic detectors will be also discussed. 

\begin{figure}[t]
  \centering
    \includegraphics[width=12cm]{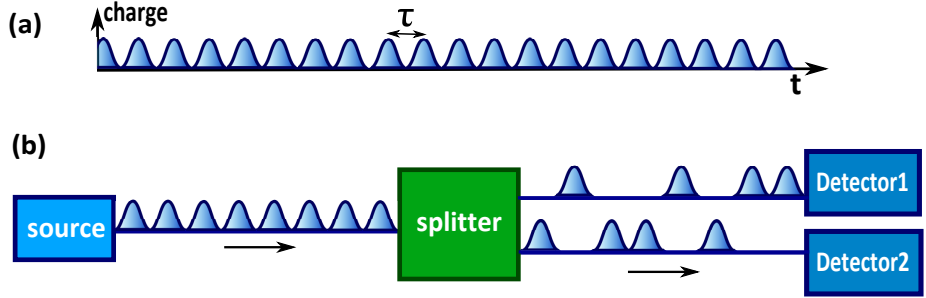}
  \caption{(a) An ordered train of wave packet. (b) Schematics of an experiment. The splitter may be quantum point contact, quantum dot, interferometer, molecular bridge, etc.}
\end{figure}

\newpage
\section{Shot Noise Power Spectrum}
\subsection{General form}
\subsubsection{Current}
First of all, we need to describe current as a train of wave packets. Suppose that the wave packet shape is $g(t)$, whose Fourier transform is $\mathscr{F}\big[g(t)\big] = \tilde{g}(f)$. $g(t)$ should be normalized as $\int g(t)dt =1$, which one can also regard as the quantum probability. The train can be expressed by the convolution of train of spikes and $g(t)$. The time difference between the spikes is $\tau$, which is at the scale of $\tau \approx \frac{h}{eV}$. The $k$th time slot appears at the time $k\tau$. The expression of the train of spikes is as follows. 
\begin{eqnarray}
h_{i}(t) & = & \sum_{k=1}^{N} \delta(t-t_{k}) \quad (t_{k} = k\tau)
\end{eqnarray}
There are $N$ slots. To describe the current after the system, we introduce the occupation of each slot, $o_{k}$ ($k=1\cdots N$).
\begin{eqnarray}
h(t) & = & \sum_{k=1}^{N}o_{k} \delta(t-t_{k})
\end{eqnarray}
Now the train of the wave packets is
\begin{eqnarray}
h(t)*g(t) & = & \sum_{k=1}^{N}o_{k} g(t-t_{k})
\end{eqnarray}
The electric current $I(t)$ has to be multiplied with charge $q$. 
\begin{eqnarray}
I(t) & = & \sum_{k=1}^{N} q o_{k} g(t-t_{k})
\end{eqnarray}
This is absolutely everything one wishes to know about an electric current.

\subsubsection{Average current and current fluctuation}
Defining $\mathscr{T}= N\tau$, the average of the impinged current is 
\begin{eqnarray}
\langle I_{i} \rangle & = & \lim_{\mathscr{T} \rightarrow \infty}\frac{q}{\mathscr{T}}\sum_{k=1}^{N}\int_{-\mathscr{T}/2}^{\mathscr{T}/2}g(t-t_{k})dt\nonumber\\
& = & \lim_{\mathscr{T} \rightarrow \infty}\frac{qN}{\mathscr{T}}\nonumber\\
& = &\frac{q}{\tau}
\end{eqnarray}
The average transmitted current is
\begin{eqnarray}
\langle I \rangle & = & \lim_{\mathscr{T} \rightarrow \infty}\frac{q}{\mathscr{T}}\sum_{k=1}^{N}o_{k}\int_{-\mathscr{T}/2}^{\mathscr{T}/2}g(t-t_{k})dt\nonumber\\
& = & \lim_{\mathscr{T} \rightarrow \infty}\frac{q\sum o_{k}}{\mathscr{T}}\nonumber\\
& = & \lim_{\mathscr{T} \rightarrow \infty}\frac{Nq}{\mathscr{T}}\frac{\sum o_{k}}{N}\nonumber\\
& = & \langle o_{k} \rangle\frac{q}{\tau}\nonumber\\
& = & T\langle I_{i} \rangle
\end{eqnarray}
The shot noise is the specral density of the current fluctuation. The fluctuation in the occupations is given by $\delta o_{k}=o_{k}- \langle o_{k} \rangle$.
\begin{eqnarray}
\delta I(t) & = & \sum_{k=1}^{N} q\delta o_{k}g(t-t_{k})
\end{eqnarray}

\subsubsection{Correlation function}
Now, let us calculate the correlation function.
\begin{eqnarray}
C(t) & = & \lim_{\mathscr{T} \rightarrow \infty} \frac{1}{\mathscr{T}}\int_{-\mathscr{T}/2}^{\mathscr{T}/2} dt' \delta I(t')\delta I(t'+t)\nonumber\\
& = & \lim_{\mathscr{T} \rightarrow \infty} \frac{q^2}{\mathscr{T}} \sum_{k,k'} \delta o_{k}\delta o_{k'} \int_{-\mathscr{T}/2}^{\mathscr{T}/2} dt' g(t'-t_{k})g(t'+t-t_{k'})\nonumber\\
& = & \lim_{\mathscr{T} \rightarrow \infty} \frac{q^2}{\mathscr{T}} \sum_{k,k'} \delta o_{k}\delta o_{k'} \int dt'g(t')g(t+t'+t_{k}-t_{k'})\nonumber\\
& = & \lim_{\mathscr{T} \rightarrow \infty} \frac{q^2}{\mathscr{T}} \sum_{m=-(N-1)}^{N-1} \sum_{n=[1:N-m], m\ge 0} ^{n=[|m|+1:N], m<0}  \delta o_{n+m} \delta  o_{n}  \int dt'g(t')g(t+t'+t_{m+n}-t_{n})\nonumber\\
& = & \lim_{\mathscr{T} \rightarrow \infty} \frac{q^2}{\mathscr{T}} \sum_{m=-(N-1)}^{N-1} \sum_{n=[1:N-m], m\ge 0} ^{n=[|m|+1:N], m<0} \delta o_{n+m} \delta  o_{n}  \int dt'g(t')g(t+t'+m\tau)\nonumber\\
& = & \lim_{\mathscr{T} \rightarrow \infty} \frac{q^2}{\mathscr{T}} \sum_{m=-(N-1)}^{N-1} (N-|m|) \sum_{n=[1:N-m], m\ge 0} ^{n=[|m|+1:N], m<0} \frac{\delta o_{n+m} \delta  o_{n}}{N-|m|}  \int dt'g(t')g(t+t'+m\tau)\nonumber\\
& = & \lim_{\mathscr{T} \rightarrow \infty} \frac{Nq^2}{\mathscr{T}} \sum_{m=-(N-1)}^{N-1} \Big(1-\frac{|m|}{N}\Big) \langle \delta o_{n+m} \delta  o_{n} \rangle_{N-|m|} G(t+m\tau)\nonumber\\
& \approx & \lim_{\mathscr{T} \rightarrow \infty} \frac{Nq^2}{\mathscr{T}} \sum_{m}^{*} \langle \delta o_{n+m} \delta  o_{n} \rangle G(t+m\tau)\nonumber\\
& = & \frac{q^2}{\tau} \sum_{m}^{*} \langle \delta o_{n+m} \delta  o_{n} \rangle G(t+m\tau)\nonumber\\
& = & q\langle I_{i} \rangle \sum_{m}^{*} \langle \delta o_{n+m} \delta  o_{n} \rangle G(t+m\tau)
\end{eqnarray}
$\langle \delta o_{n+m} \delta  o_{n} \rangle_{N-|m|}$ is the average of $\delta o_{n+m} \delta  o_{n}$ over $N-|m|$ events. The approximation from the 7th to the 8th line was done with one assumption. Namely, there are finite correlations, if at all, only among small $m$ ($m<<N$). Certainly, it is senseless that a tunneling event affects 1000 events later. Hence, $\langle \delta o_{n+m} \delta  o_{n} \rangle_{N-|m|} = \langle \delta o_{n+m} \delta  o_{n} \rangle$ in the large $N$ limit. Therefore, the sum over $m$ implicitly means summing over some small numbers (say, up to $m \approx 10$). We denote it as $\sum_{m}^{*}$ without specifying the upper bound for $m$. Besides, we have defined the correlation function of $g(t)$ as $G(t)$.
\begin{eqnarray}
G(t) & = & \int dt'g(t')g(t'+t) \quad, \quad \tilde{G}(f) \quad = \quad |\tilde{g}(f)|^{2}
\end{eqnarray}

\subsubsection{Power spectrum}
The shot noise power spectrum is $S(f) =  \bar{S}(f) + \bar{S}(-f)$. Combining $f$ and $-f$ makes it enough to take into account only $m\ge 0$ component.
\begin{eqnarray}
S(f) & = & q\langle I_{i} \rangle \sum_{m\ge 0}^{*} \langle \delta o_{n+m} \delta o_{n} \rangle (e^{-im\omega\tau}+e^{im\omega\tau})\tilde{G}(f)\nonumber\\
& = & 2q \langle I_{i} \rangle \sum_{m\ge 0}^{*} \langle \delta o_{n+m} \delta o_{n} \rangle cos(m\omega\tau)\tilde{G}(f)\nonumber\\
& = & 2q \langle o_{n} \rangle \langle I_{i} \rangle \sum_{m\ge 0}^{*} \frac{\langle \delta o_{n+m} \delta o_{n} \rangle}{\langle o_{n} \rangle}cos(m\omega\tau)\tilde{G}(f)\nonumber\\
& = & 2q\langle o_{n} \rangle F(f) \tilde{G}(f) \langle I_{i}\rangle
\end{eqnarray}
Thus, the shot noise power spectrum is proportional to the charge $q$ and the average transmitted current $\langle I\rangle$ and the frequency profile is embedded in $\tilde{G}(f)$ and $F(f)$. The shape of the wave packets only comes in $\tilde{G}(f)$ whose bandwidth is the order of $\tau^{-1}$. The Fano factor contains the information of the correlation among the tunneling events.

\subsubsection{Absence of white noise in ordered currents}
When $o_{k} = 1$, the current is an ordered train of wave packets with no fluctuatioin, namely $\delta o_{k} = 0$. Such ordered current is a signal with periodicity of $\tau$. Thus, power spectrum of such current has a disrete spectrum being finite only at the harmonics of $\tau^{-1}$. Hence, the white shot noise vanishes. In general, if $o_{k}$ changes in any periodic manner with a periodicity $p$, its power spectrum is finite only at the harmonics of $p^{-1}$. Any periodic (ordered) currents has no shot noise.

\subsection{Wave packet shape}
\subsubsection{Case of spikes}
The simplest example of wave packet is spikes, delta function $g(t) = \delta (t-t_{k})$. The Fourier transform, 
autocorrelation function, power spectrum, and resultant shot noise are
\begin{eqnarray}
\mathscr{F}[\delta (t-t_{k})] & = & 1\\
G(t) & = &  \delta (t)\\
\tilde{G}(f) & = & 1 \\
S(f) & = & 2qF(f) \langle I \rangle
\end{eqnarray}
For the train of spikes, the power spectrum is white noise (suppose $F(f)=$ constant) up to infinite frequency and only the Fano factor gives a frequency dependence. Of course, the infinitely sharp wave packet is not physical because one needs infinitely high energy to localize a wave into infinitely sharp. Regardless these concerns, it still captures the essence.

\subsubsection{Case of square wave packets}
We consider the case of square wave packets of duration $\eta$ and height $1/\eta$. $g(t)=\frac{1}{\eta}$ for $|t|\le \frac{\eta}{2}$ and $0$ for $|t| > \frac{\eta}{2})$. The Fourier transform, autocorrelation function, power spectrum, and shot noise are
\begin{eqnarray}
\mathscr{F}[g(t)] & = & \frac{sin(\pi f\eta)}{\pi f\eta}\\
G(t) & = & \frac{1}{\eta}\Big[1-\frac{|t|}{\eta}\Big] \quad (|t|\le \eta), \quad 0 \quad (|t| > \eta)\\
\tilde{G}(f) & = & \Big[\frac{sin(\pi f\eta)}{\pi f\eta} \Big]^{2}\\
S(f) & = & 2qF(f)\Big[\frac{sin(\pi f\eta)}{\pi f\eta} \Big]^{2} \langle I \rangle
\end{eqnarray}
For general $\eta$, the shot noise exhibits a white spectrum. Here, note that the case of $\eta=\tau$ means a strictly DC current withouth any gap between the wave packets. Regardig the section about the absence of shot noise in periodic signals, one can also understand it because the sinc function becomes zero at every harmonics of frequency $\frac{1}{\tau}$.

\subsubsection{Case of sinc wave packet}
We consider the case of sinc wave packets $g(t)=f_{0}\frac{sin(\pi f_{0}t)}{\pi f_{0}t}$. The Fourier transform, 
autocorrelation function, power spectrum, and shot noise are
\begin{eqnarray}
\mathscr{F}[g(t)] & = & 1 \quad (|f|\le \frac{f_{0}}{2}), \quad 0 \quad (|f| > \frac{f_{0}}{2})\\
G(t) & = & \frac{sin(\pi f_{0}t)}{\pi f_{0}t}\\
\tilde{G}(f) & = &  1 \quad (|f|\le \frac{f_{0}}{2}), \quad 0 \quad (|f| > \frac{f_{0}}{2})\\
S(f) & = & 2qF(f)\tilde{G}(f) \langle I \rangle
\end{eqnarray}
In the limit of large $f_{0}$, this is reduced to the case of spikes. Though it was presented here, it is not very physical actually. Since there is no negative probability in quantum wave packets. (The sinc function takes negative value too.) In many transport situations, a DC bias $eV$ is applied to the system above the Fermi energy. One can view the electrons within the energy $eV$ form wave packets (in the small enough bais range, one can assume the dispersion of electrons is linear $\omega = v_{F}k$, where $v_{F}$ is the Fermi velocity). $h/eV$ is the time scale of inter-electron time difference. Hence, $hv_{F}/eV$ is more or less the electron's wave packet size. The power spetrum of such a current should have a high-frequency cut at the frequency $eV/h$ as the highest energy of the system. The sinc wave packets generates a rectangle power spectrum whose cut off is at $f=\frac{f_{0}}{2}$. This gives an insight to the actual shape of the electron wave packets.

\subsubsection{Case of Gaussian wave packet}
We consider the case of a Gaussian wave packet $g(t)=\frac{1}{\sqrt{2\pi}u}e^{-\frac{t^2}{2u^2}}$. The Fourier transform, autocorrelation function, power spectrum, and shot noise are
\begin{eqnarray}
\mathscr{F}[g(t)] & = & \sqrt{\frac{\pi}{a}}e^{-\frac{\omega^2}{2a}}\\
G(t) & = & \frac{1}{2\sqrt{\pi}u} e^{-\frac{t^2}{4u^2}}\\
\tilde{G}(f) & = & e^{-(\omega u)^2}\\
S(f) & = & 2qF(f)e^{-(\omega u)^2} \langle I \rangle
\end{eqnarray}
Gaussian wave packets are easy to handle and often employed. However, note that it has finite, yet vanishing, spectrum even beyond $eV/h$.

We have seen several examples of the wave packet shapes. Each shape generates a unique power spectrum. However, there is one common feature. As far as the integral of the wave packet over time is unity ($\int dt g(t) =1$), $\tilde{G}(f)$ goes to unity in the limit of low frequency, $\tilde{G}(f) \rightarrow 1$. This is trivial because integrating over time is nothing but taking only the DC component. In the next section, we will see one example that $\tilde{G}(f)$ lacks the low-frequency component.

\subsubsection{Case of paired wave packets}
Next, consider two wave packets $g'(t)$ which are temporally separated by $\eta$ and have opposite sign of charge, $q$ and $-q$, whose sum is zero. Thus, this is an uncoventional situation, where finite shot noise without net current exists. 
\begin{eqnarray}
g(t) & = & g'(t) - g'(t+\eta)
\end{eqnarray}
The autocorrelation of $g(t)$ and its fourier spectrum is 
\begin{eqnarray}
G(t) & = & \int dt'\Big[g'(t) - g'(t+\eta)\Big]\Big[g'(t+t') - g'(t+t'+\eta)\Big] \nonumber\\
& = & \int dt'
 \Big[g'(t')g'(t+t')+g'(t'+\eta)g'(t+t'+\eta)\\
&& \quad \quad \quad \quad \quad -g'(t')g'(t+t'+\eta)-g'(t'+\eta)g'(t+t') \Big]\nonumber\\
& = & 2G'(t)-G'(t+\eta)-G'(t-\eta)\\
\tilde{G}(f) & = & \Big[2 - e^{-i \omega \eta}- e^{i \omega \eta}\Big]\tilde{G}'(f)\quad = \quad 4sin^{2}(\omega \eta)\tilde{G}'(f)
\end{eqnarray}
Though the net current is zero ($\langle I_{i} \rangle=\langle I \rangle=0$), we can define the particle (pair) number current such that
\begin{eqnarray}
\langle I_{p,i} \rangle & = & \lim_{\mathscr{T} \rightarrow \infty}\frac{N}{\mathscr{T}} \quad = \quad \frac{1}{\tau}\\
\langle I_{p} \rangle  & = & \lim_{\mathscr{T} \rightarrow \infty}\frac{\langle o_{k} \rangle N}{\mathscr{T}} \quad = \quad \frac{1}{\tau} \quad = \quad \frac{\langle o_{k} \rangle}{\tau}.
\end{eqnarray}
Using this, shot noise power spectrum of random train of the pair is as the following.
\begin{eqnarray}
S(f) & = & 8q^{2}\langle I_{p} \rangle F(f) sin^{2}(\omega \eta)\tilde{G}'(f)
\end{eqnarray}
The low-frequency component of the shot noise is vanishingly small. Although this example may sound unphysical, there is a model system that behaves as this, which is a capacitively-coupled 1D conductors. Generating a wideband power in a conductor capacitively induces a high-frequency power with a low-frequency cutoff depending on the resistances and the capacitance of the counductors in the other conductor. The resultant spectrum of the other conductor resembles the power spectrum given above.

\subsection{Fano factor and effective charge}
\subsubsection{Case of Poissonian wave packets}
So far, we have not discussed $o_{n}$ in detail. The ordered train of wave packets is partitioned at the splitter and becoems a stochastic train of them. Quantities such as $\langle o_{n} \rangle$ and $\langle o_{n}^{2} \rangle$ can be calculated once a specific distribution is given. Let us consider the simplest example, Poissonian distribution. For a Poissonian process, the mean and variance are the same.
\begin{eqnarray}
\langle \delta o_{n}^{2} \rangle & = & \langle o_{n} \rangle \quad \Rightarrow \quad F(f) \quad = \quad 1\\
S(f) & = & 2q\langle o_{n} \rangle\langle I_{i} \rangle\tilde{G}(f) \\
S(f\approx 0) & = & 2q\langle I \rangle
\end{eqnarray}
Here, we used the fact that $\tilde{G}(f\approx 0)\approx 1$ at low frequencies $f << \tau^{-1}$ and $\langle I \rangle = \langle o_{n} \rangle\langle I_{i} \rangle$. The Schottkey formula for shot noise is obtained. 

\subsubsection{Case of independent Fermionic wave packets}
Here, we assume that those wave packets follow the Fermi statistics. The impinged current is ordered such that $o_{n}=1$. The tunneling process is the binomial distribution. The wave packets are transmitted (reflected) with a probability $T$ ($1-T=R$) at each tunneling event. We assume there is no correlation among the events.
\begin{eqnarray}
(T+R)^{N} & = & \sum_{n=0}^{N} \frac{N!}{(N-n)!n!} T^{n}R^{N-n}
\end{eqnarray}
, where $n$ is the number of transmitted wave packets. Given the distribution, the expectation values of $n$ and $n^2$ are
\begin{eqnarray}
\langle n \rangle & = & T\frac{\partial}{\partial T}[T+R]^{N}\quad = \quad NT\\
\langle n^{2} \rangle & = &  T\frac{\partial}{\partial T}T\frac{\partial}{\partial T}[T+R]^{N}\quad = \quad (NT)^{2}+NT(1-T)
\end{eqnarray} 
Thus, the average and variance of the occupation numbers are 
\begin{eqnarray}
\langle o_{n} \rangle  & = & \frac{\langle n \rangle}{N} \quad = \quad T\\
\langle \delta o_{n}^{2} \rangle & = & \frac{\langle n^{2} \rangle-\langle n \rangle^{2}}{N} \quad = \quad T(1-T)\\
F(f) & = & 1-T \quad \equiv \quad F_{if}
\end{eqnarray} 
The subscript "$if$" stands for the independent Fermionic case. The Fano factor is less than unity, which is smaller than the Poissonian case. Hence, sometimes this is called subpoissonian shot noise. Just to show the full expression, the shot noise of the independent Fermionic wave packets is
\begin{eqnarray}
S(f) & = & 2q\langle o_{n} \rangle F_{0}\langle I_{i} \rangle \tilde{G}(f) \quad = \quad 2qT(1-T)\tilde{G}(f)\langle I_{i} \rangle
\end{eqnarray}
Let us denote the shot noise for the independent fermionic case as $S_{if}(f)$.
\begin{eqnarray}
S_{if}(f) & = & 2qTF_{if}\tilde{G}(f) \langle I_{i} \rangle
\end{eqnarray}

\subsubsection{Decomposition of Fano factor}
Let us decompose the Fano factor to make its structure clearer.
\begin{eqnarray}
F(f) & \equiv & \frac{\langle \delta o_{n}^{2}\rangle}{\langle o_{n} \rangle}+\sum_{m>\ge}^{*} \frac{\langle \delta o_{n+m} \delta o_{n} \rangle}{\langle o_{n} \rangle}cos(m\omega\tau) \nonumber\\
& = & \frac{\langle \delta o_{n}^{2}\rangle}{\langle o_{n} \rangle} + \sum_{m>0}^{*} \frac{\langle \delta o_{n+m} \delta o_{n} \rangle}{\langle o_{n} \rangle}cos(m\omega\tau)\nonumber\\
& = & \frac{\langle \delta o_{n}^{2}\rangle}{\langle o_{n} \rangle}\Big[1 + \sum_{m>0}^{*} \frac{\langle \delta o_{n+m} \delta o_{n} \rangle}{\langle \delta o_{n}^{2} \rangle}cos(m\omega\tau) \Big]\nonumber\\
& = & F_{0}\Big[1 + \sum_{m>0}^{*} \frac{\langle \delta o_{n+m} \delta o_{n} \rangle}{\langle  \delta o_{n}^{2}\rangle}cos(m\omega\tau) \Big]\nonumber\\
& = & F_{0}\Lambda(f) \quad (f\ge 0)
\end{eqnarray}
The first term describes the correlation within an event and the second term describes the correlation among events. Let us simply say that the first and second term corresponds to the intra-event and inter-event correlation, respectively. While the intra-event correlation gives a whilte spectrum for the Fano factor, the inter-event correlation induces the frequency dependence. 

\subsubsection{Low frequency shot noise and effective charge}
At low frequencies $f << \tau^{-1}$, it is clear that the low-frequency component of the shot noise is insenstive to the details of the wave packet shape as long as the $\int dtg(t)=1$, namely, its DC component is unity. Furthermore, $cos(m\omega\tau)$ that comes into the Fano factor goes to 1 in the low frequency limit. In short, $cos(m\omega\tau) \rightarrow 1$ and $\tilde{G}(f) \rightarrow 1$ in the limit of $f \rightarrow 0$.
\begin{eqnarray}
\Lambda(f \rightarrow 0) & = & \Big[1 + \sum_{m>0}^{*} \frac{\langle \delta o_{n+m} \delta o_{n} \rangle}{\langle \delta o_{n}^{2}\rangle}\Big] \quad \equiv \quad \Lambda_{0}
\end{eqnarray}
Note that $\Lambda_{0}=1$ in the independent Fermionic case. Therefore, the low-frequency shot noise is as follows.
\begin{eqnarray}
S(f \rightarrow 0) & = & 2qTF_{o}\Lambda_{0}\langle I_{i}\rangle 
\end{eqnarray}
Now we are at the stage to introduce the concept of effective charge. The effective charge is a covenient picture to capture the tunneling process intuitively instead of talking about the exact values of Fano factor. It interpretes the tunneling process such that Fermionic wave packets with charge $q^{*}$ independently tunnels across the partitioner. Namely,
\begin{eqnarray}
S(f \rightarrow 0) & \equiv & 2q^{*}TF_{if}\langle I_{i}\rangle 
\end{eqnarray}
This is the definion of the effective charge. We can rewrite it as follows.
\begin{eqnarray}
q^{*} & = & \frac{S(f \rightarrow 0)}{S_{if}(f \rightarrow 0)}\cdot q \quad = \quad \frac{F_{0}\Lambda_{0}}{F_{if}}\cdot q
\end{eqnarray}
As mentioned before, the effective charge is effective since the details of wave packet shape is irrelevant in the low-frequency limit. There are two ways to modify the effective charge $q^{*}$ from $q$, which are intra- and inter-events correlations. To look at the effect of $F_{0}$ and $\Lambda_{0}$ separately, we will consider two cases, $F_{0}=F_{if}$ and $\Lambda_{0}=1$ in details.

\subsubsection{Effective charge via correlation among tunnelings}
In the independent Fermionic case, the charge of the tunneling entity cannot be deviate from $q$. We assume here, for simplicity, that $F_{0}=F_{if}$, which corresponds to systems with mere electrons as only available excitations (no exotic fractional quasiparticles). Thus, the effective charge is now 
\begin{eqnarray}
q^{*} & = & \Lambda_{0} q
\end{eqnarray}
Let us take several examples to capture the idea of inter-event correlation. Note that the present method does not provide the microscopic mechanism how such correlation is induced. It rather allows to compute the effective charge for a given set of correlations. Hence, one can also come up with a list of possible correlations to yield the observed effective charge. To resolve among the possible types of inter-event correlations, one has to measure the full power spectrum. Suppose there is a correlation in only neighboring events and no further correlation.
\begin{eqnarray}
\langle \delta o_{n+1} \delta o_{n} \rangle & = & \frac{2}{3}\langle \delta o_{n}^{2}\rangle \\
\rightarrow \quad \Lambda_{0} & = & 1 + \frac{2}{3} \quad = \quad \frac{5}{3}
\end{eqnarray}
There are two main processes where one wave packet transmits alone and two consecutive wave packets transmit together. The two-particle process looks as if they are bunching. Sometimes it is called particle bunching. The effective charge is $\frac{5}{3}q$. Next, suppose there is an anti-correlation in only neighboring events.
\begin{eqnarray}
\langle \delta o_{n+1} \delta o_{n} \rangle & = & -\frac{2}{3}\langle \delta o_{n}^{2}\rangle \\
\rightarrow \quad \Lambda_{0}  & = & 1 - \frac{2}{3} \quad = \quad  \frac{1}{3}
\end{eqnarray}
Thus, the effective charge is $\frac{1}{3}q$. Let us look at another example with a further inter-event correlation.
\begin{eqnarray}
\langle \delta o_{n+2} \delta o_{n} \rangle & = & \frac{1}{3}\langle \delta o_{n}^{2}\rangle\\
\langle \delta o_{n+1} \delta o_{n} \rangle & = & \frac{2}{3}\langle \delta o_{n}^{2} \rangle\\
\rightarrow \quad \Lambda_{0}  & = & 1 + \frac{2}{3} + \frac{1}{3} \quad = \quad  2
\end{eqnarray}
The effective charge is $2q$. The last example also gives rise to the effective charge of $2q$ but with different inter-event correlations.
\begin{eqnarray}
\langle \delta o_{n+3} \delta o_{n} \rangle & = & \frac{1}{6}\langle \delta o_{n}^{2}\rangle\\
\langle \delta o_{n+2} \delta o_{n} \rangle & = & \frac{1}{3}\langle \delta o_{n}^{2} \rangle\\
\langle \delta o_{n+1} \delta o_{n} \rangle & = & \frac{1}{2}\langle \delta o_{n}^{2} \rangle\\
\rightarrow \quad \Lambda_{0} & = & 1 + \frac{1}{2} + \frac{1}{3} + \frac{1}{6} \quad = \quad 2
\end{eqnarray}
In actual experiments, one often faces irrational effective charges even in systems with, presumably, mere electrons. Quantum dots are good example. With multiple states within the bias window or cotunneling by higher order processes can enhance the effective charge from mere $q$. In order to reveal whether the effective charge is the elementary charge of the system or not, one should take $T \approx 0$ or $T \approx 1$ where the single particle process dominates and the distribution reduces to the Poissonian case.

\subsubsection{Effective charge via uncorrelated multiple occupation}
Suppose now each slot of an ordered train accomodates up to $K$ wave packets ($k$ = 0, 1, 2, ..., $K$) and there is no inter-event correlation ($\Lambda_{0}=1$). 
\begin{eqnarray}
q^{*} & = & \frac{F_{0}}{F_{if}}\cdot q
\end{eqnarray}
Let us denote the probability that $k$ wave packets transmit together as $T_{k}$ with $\sum_{k=0}^{K}T_{k} = 1$. The distribution that we employ now is the multinomial distribution. There are $NK$ wave packets over the train and $n_{k}$ tunneling events that $k$ wave packets transmit together. Hence, $\sum_{k=0}^{K}n_{k}=N$. 
\begin{eqnarray}
\Bigg[\sum_{k=0}^{K}T_{k} \Bigg]^{N} & = & \sum_{(n_{0}, n_{1}, \cdots, n_{K})}\frac{N!}{n_{0}!n_{1}!\cdots n_{K}!} T_{0}^{n_{0}}T_{1}^{n_{1}}\cdots T_{K}^{n_{K}}
\end{eqnarray}
Given the distribution, the expectation values of $n_{k}$ and $n_{k}^2$ are
\begin{eqnarray}
\langle n_{k} \rangle & = & T_{k}\frac{\partial}{\partial T_{k}}\Bigg[ \sum_{s=0}^{K}T_{k} \Bigg]^{N}\quad = \quad NT_{k}\\
\langle n_{k}^{2} \rangle & = &  T_{k}\frac{\partial}{\partial T_{k}}T_{k}\frac{\partial}{\partial T_{k}}\Bigg[ \sum_{s=0}^{K}T_{s} \Bigg]^{N}\quad = \quad (NT_{k})^{2}+NT_{k}(1-T_{k})
\end{eqnarray}
The expectation values of the mean and variance of the occupation $o_{n}$ and the Fano factor read
\begin{eqnarray}
\langle o_{n} \rangle & = & \frac{1}{N} \sum_{r=0}^{K} k\langle n_{k} \rangle \quad = \quad \sum_{k=0}^{K} kT_{k}\\
\langle \delta o_{n}^{2} \rangle & = & \frac{1}{N} \sum_{k=0}^{K} k^{2}\langle \delta n_{k}^{2} \rangle \quad = \quad \sum_{k=0}^{K} k^{2} T_{k}(1-T_{k})\\
F_{0} & = & \frac{\sum_{k=0}^{K} k^{2} T_{k}(1-T_{k})}{\sum_{k=0}^{K} kT_{k}}
\end{eqnarray}
One may define the normalized transmission probability $T_{nrm}$ and the normalized Fano factor as
\begin{eqnarray}
T_{nrm} & = & \frac{\langle o_{n} \rangle}{K} \quad = \quad  \sum_{k=0}^{K} \frac{k}{K}T_{k}\\
F_{0,nrm} & = & \frac{F_{0}}{K} \quad = \quad \frac{\sum_{k=0}^{K} \big(\frac{k}{K}\big)^{2} T_{k}(1-T_{k})}{\sum_{k=0}^{K} \frac{k}{K}T_{k}}
\end{eqnarray}
$T_{nrm}$ is nothing but the transmission that is measured in actual conductance measurements. Note that the initial current is $\langle I_{i} \rangle = \frac{Kq}{\tau}$ and the number of events per unit time is $\frac{1}{\tau}$.
\begin{eqnarray}
S(f) & = & 2q\langle o_{n} \rangle F_{0} \frac{\langle I_{i} \rangle}{K} \tilde{G}(f)\quad = \quad 2KqT_{nrm}F_{0,nrm}\langle I_{i} \rangle \tilde{G}(f)
\end{eqnarray}
Alhough the multiple occupation case sounds unphsycial, the edge state of fractional qunatum Hall effect, e.g. the filling factor 1/3, may be modeled as the case of $K=3$ with $q=\frac{e}{3}$. Let us consider the low-frequency part of the shot noise given in the last section. $S(f\approx 0) = 2KqT_{nrm}F_{0,nrm}\langle I_{i} \rangle$. The effective charge can be defined with the shot noise of the independent Fermionic wave packets with transmissioni probability $T_{nrm}$, which is $2qT_{nrm}(1-T_{nrm})\langle I_{i} \rangle$.
\begin{eqnarray}
q^{*} & = & \frac{F_{0}}{F_{if,nrm}}\cdot q \quad = \quad \frac{KF_{0,nrm}}{F_{if,nrm}}\cdot q
\end{eqnarray}
,where $F_{if,nrm}=1-T_{nrm}$ and $F_{0}=KF_{0,nrm}$.

\subsubsection{Effective charge via correlated multiple occupation}
Though we do not go into details, for the completion, we show the expression of the effective charge in the case of multiple occupation with correlation among the tunnelings.
\begin{eqnarray}
q^{*} & = & \frac{KF_{0,nrm}\Lambda_{0}}{F_{if,nrm}}\cdot q
\end{eqnarray}

\subsubsection{Case of uncorrelated double occupation}
Here, we consider the case of $K=2$ explicitly.\\
\noindent(1):\quad $(T_{0},T_{1},T_{2})=(R, T, 0)$\\
$(\langle o_{n} \rangle, \langle \delta o_{n}^{2} \rangle, T_{nrm}, F_{0}) \quad = \quad (T,\quad T(1-T),\quad \frac{T}{2},\quad 1-T)$
\begin{eqnarray}
q^{*}(T) & = & \frac{1-T}{1-\frac{T}{2}}q ,\quad q^{*}(T \rightarrow 0) \quad = \quad q
\end{eqnarray}
\noindent(2):\quad $(T_{0},T_{1},T_{2})=(R, \frac{T}{2}, \frac{T}{2})$\\
$(\langle o_{n} \rangle, \langle \delta o_{n}^{2} \rangle, T_{nrm}, F_{0}) \quad = \quad (3\frac{T}{2},\quad  5\frac{T}{2}(1-\frac{T}{2}),\quad \frac{3}{2}\frac{T}{2},\quad \frac{5}{3}(1-\frac{T}{2}))$
\begin{eqnarray}
q^{*}(T) & = & \frac{5}{3}\frac{1-\frac{T}{2}}{1-\frac{3T}{4}}q, \quad q^{*}(T \rightarrow 0) \quad = \quad \frac{5}{3}q
\end{eqnarray}
\noindent(3):\quad $(T_{0},T_{1},T_{2})=(R, 0, T)$\\
$(\langle o_{n} \rangle, \langle \delta o_{n}^{2} \rangle, T_{nrm}, F_{0}) \quad = \quad (2T,\quad 4T(1-T),\quad  T,\quad 2(1-T))$
\begin{eqnarray}
q^{*}(T) & = & 2q, \quad q^{*}(T \rightarrow 0) \quad = \quad 2q 
\end{eqnarray}
\noindent(4):\quad $(T_{0},T_{1},T_{2})=(R, aT, (1-a)T)$, $\quad \quad (0\leq a \leq 1)$\\
$(\langle o_{n} \rangle, \langle \delta o_{n}^{2} \rangle) = ((2-a)T, \quad (4-3a)T-(4-8a+5a^{2})T^{2})$\\
Here, we look at them a bit more carefully.
\begin{eqnarray}
T_{nrm} & = & \frac{1}{2}(2-a)T\\
F_{0} & = & \frac{4-3a}{2-a}\Big(1-\frac{4-8a+5a^{2}}{4-3a}T\Big)\\
q^{*}(T,a) & = & \frac{4-3a}{2-a}\cdot\frac{1-\frac{4-8a+5a^{2}}{4-3a}T}{1-\frac{1}{2}(2-a)T}\cdot q \\
q^{*}(T \rightarrow 0,a) & \rightarrow & \frac{4-3a}{2-a}q
\end{eqnarray}
It is interesting to notice that $a=\frac{2}{3}$ yields $q^{*}=\frac{1}{2}(q+2q)=\frac{3}{2}q$. The equal mixing of the one-particle and two-particle tunneling events ($a=\frac{1}{2}$) does not simply leads to the effective charge right at the mid value $q^{*}=\frac{3}{2}q$. This is because the variance goes as quadratic in the tunneling charge.

\subsubsection{Case of uncorrelated triple occupation}
Here, we consider the case of $K=3$ explicitly. 
\noindent(1):\quad $(T_{0},T_{1},T_{2},T_{3})=(R, aT, bT, (1-a-b)T)$\\
$(\langle o_{n} \rangle, \langle \delta o_{n}^{2} \rangle, T_{nrm}, F_{0}) \quad = \quad (T,\quad T(1-T),\quad  \frac{1}{3}T,\quad (1-T))$
\begin{eqnarray}
T_{nrm} & = & \frac{1}{3}(3-2a-b)T\\
F_{0} & = & \frac{(9-8a-5b)-(9+10a^{2}+13b^{2}-9a-9b-9ab)T}{3-2a-b}\\
q^{*}(T,a,b) & = & \frac{9-8a-5b}{3-2a-b}\cdot\frac{1-\frac{(9+10a^{2}+13b^{2}-9a-9b-9ab)}{9-8a-5b}T}{1-\frac{1}{3}(3-2a-b)T}\cdot q\\
q^{*}(T\rightarrow 0,a,b) & = & \frac{9-8a-5b}{3-2a-b}\cdot q 
\end{eqnarray}

\subsubsection{Case of multiple independent Fermionic conductors}
Suppose now there are multiple (1D) conductors participating in the transport. However, to keep it simple, let us assume they are just independent from each other (no wave packet swapping). Let us index the conductors as $l=1,2,\cdots,L$. The full expression is quite straightforward. If we suppose all the wave packet shapes in the conductors are identical, $G_{l}(f) = G_{f}$.
\begin{eqnarray}
S(f) & = & \sum_{l=1}^{L} 2q\langle o_{n,l} \rangle F_{0,l}\langle I_{i} \rangle \tilde{G}_{l}(f)\\
& = & \sum_{l=1}^{L}2qT_{l}(1-T_{l})\langle I_{i} \rangle \tilde{G}_{l}(f)\\
& = & \Big[\sum_{l=1}^{L}2qT_{l}(1-T_{l})\Big]\langle I_{i} \rangle \tilde{G}(f)
\end{eqnarray}
One can think of a model system as a quantum dot or single molecule bridge where multiple channels for the transmission can be available. For that a more careful consideration such as assigning different $G_{l}(f)$ and $\langle I_{i,l} \rangle$. The index $l$ is for each transport channel. In the Landauer's paper, he considered the shot noise as a function of the DC bias $eV$. First, he split this energy window into many tiny energy segments $\Delta E$, calculated the shot noise, and then summed over the entire energy with occupations follwoing the Fermi-Dirac distribution. Regarding each segment as a conductor, we can recover and even extend the Landauer's approach with the arguments given so far.

\subsubsection{Case of fluctuating tunnel barrier}
We again consider the independent Fermionic case. However, now the tunneling probability is fluctuating in time. We start with the simplest case. Namely, it takes two transmission probabilities $T_{1}$ for $0\le t \le N_{1}\tau$ and $T_{2}$ for $N_{1}\tau < t \le (N_{1}+N_{2})\tau$, where $N=N_{1}+N_{2}$. Important to note that, at the end, $N_{1}$ and $N_{2}$ should also go to infinity in taking the limit of $\mathscr{T} \rightarrow \infty$ with keeping the ratio $N_{1}/N$ and $N_{2}/N$. The distribution of the transmitted wave packets is 
\begin{eqnarray}
(T_{1}+R_{1})^{N_{1}}(T_{2}+R_{2})^{N_{2}} & = & \Big[ \sum_{n_{1}=0}^{N_{1}} \frac{N_{1}!}{(N_{1}-n_{1})!n_{1}!} T_{1}^{n_{1}}R_{1}^{N_{1}-n_{1}}\Big]\\ \nonumber
&& \quad \quad \quad \times \Big[ \sum_{n_{2}=0}^{N_{2}} \frac{N_{2}!}{(N_{2}-n_{2})!n_{2}!} T_{2}^{n_{2}}R_{2}^{N_{2}-n_{2}}\Big]
\end{eqnarray}
Thus, the number of the transmitted wave packets $n$ is $n = n_{1}+n_{2}$. We would like to calculate the expectation value and the variance of $n$. First, the expectation value of $n$ is
\begin{eqnarray}
\langle n \rangle & = & \langle n_{1} \rangle + \langle n_{2} \rangle \quad = \quad N_{1}T_{1}+N_{2}T_{2}\\
\langle o_{n} \rangle & = & \frac{\langle n \rangle}{N} \quad = \quad \frac{N_{1}}{N}T_{1}+\frac{N_{2}}{N}T_{2}
\end{eqnarray}
THen, to calculate the variance, the expectation value of $n^{2}$ is
\begin{eqnarray}
\langle n^{2} \rangle & = & \langle n_{1}^{2} \rangle + 2\langle n_{1}n_{2} \rangle + \langle n_{2}^{2} \rangle \nonumber \\
 & = & \langle n_{1}^{2} \rangle + 2\langle n_{1}\rangle \langle n_{2} \rangle + \langle n_{2}^{2} \rangle \nonumber \\
\langle \delta n^{2} \rangle & = & \langle \delta n_{1}^{2} \rangle + \langle \delta n_{2}^{2} \rangle
\end{eqnarray}
The decomposition of $\langle n_{1}n_{2} \rangle$ into $\langle n_{1}\rangle \langle n_{2} \rangle$ is simply because the binomial distribution assumes the independent events.
\begin{eqnarray}
\langle \delta o_{n}^{2} \rangle & = & \frac{\langle n \rangle}{N} \quad = \quad \frac{N_{1}}{N}T_{1}(1-T_{1})+\frac{N_{2}}{N}T_{2}(1-T_{2})
\end{eqnarray}
By induction, the extension to more fluctuating case is trivial. For the case of $M$ tunneling probabilities, the distribution function, the expectation value, and the variance is 
\begin{eqnarray}
\prod_{m=1}^{M} (T_{m}+R_{m})^{N_{m}} & = & \prod_{m=1}^{M} \Big[ \sum_{n_{m}=0}^{N_{m}} \frac{N_{m}!}{(N_{m}-n_{m})!n_{m}!} T_{m}^{n_{m}}R_{m}^{N_{m}-n_{m}}\Big]\\
\langle o_{n} \rangle & = & \frac{1}{N}\sum_{m=1}^{M}N_{m}T_{m}\\
\langle \delta o_{n}^{2} \rangle & = & \frac{1}{N} \sum_{m=1}^{M} N_{m}T_{m}(1-T_{m})
\end{eqnarray}
Now, we can take two ways in taking $\mathscr{T} \rightarrow \infty$. One is to keep he ratio $N_{m}/N$ and the other is to take $M$ to the infinity. For the former, the result given above is enough. However, for the latter, we need to know the distribution of $T_{m}$ to obtain the proper expectation values.


\newpage

\section{Practical Implementation}
\subsection{Finite Bandwidth Detector}
\subsubsection{Center frequency = DC}
In practice, the detector has a finite time resolution, $\Delta t$, with a specific shape of time-window, $Z(t)$. $Z(t)$ monitors the current reaching the detector, $I(t)$ and smears the original singal. The singal after the detector, $I_{D}(t)$, becomes the convolution of the original signal and the time window, $I(t)*Z(t)$. In the Fourier space, the detector acts as a filter with a frequency response $\tilde{Z}(f)=\mathscr{F}[Z(t)]$ centered at $f=0$Hz and a bandwidth $\Delta f$. 
\begin{eqnarray}
I_{D}(t) & = & I(t)*Z(t)\\
\mathscr{F}[Z(t)] & = & \tilde{Z}(f)
\end{eqnarray}
In most electronic systems, $\Delta t \approx (\Delta f)^{-1}$ and $\Delta t >> \tau$. Of course, the exact relation depends on the detail shape of the filter. 

Given the realistic detector, what is the measured shot noise power spectrum, $S_{D}(f)$?
\begin{eqnarray}
S_{D}(f) & = & \mathscr{F}[C_{ID}(t)]+c.c. \\
C_{D}(t) & = & \langle \delta I_{D}(t') \delta I_{D}(t'+t) \rangle\\
\delta I_{D}(t') & = & \int dt'' Z(t'')\delta I(t'-t'')\\
\delta I(t'-t'') & = & \sum_{k} o_{k} g(t'-t''-t_{k})
\end{eqnarray}
Using above fomula, let us calculate the correlation function and the power spectrum explicitly.
\begin{eqnarray}
C_{D}(t) & = & \lim_{\mathscr{T} \rightarrow \infty} \frac{1}{\mathscr{T}} \int dt' \delta I_{D}(t') \delta I_{D}(t'+t) \rangle\nonumber\\
& = & \lim_{\mathscr{T} \rightarrow \infty} \frac{q^{2}}{\mathscr{T}} \sum_{k,k'} o_{k}o_{k'} \int dt'\int dt''_{1} Z(dt''_{1})g(t'-t''_{1}-t_{k}) \int dt''_{2} Z(dt''_{2})g(t+t'-t''_{2}-t_{k'}) \nonumber\\
& = & \lim_{\mathscr{T} \rightarrow \infty} \frac{q^{2}}{\mathscr{T}} \sum_{k,k'} o_{k}o_{k'} \int dt'\int dt''_{1} Z(t''_{1})g(t'-t''_{1}-t_{k}) \int dt''_{2} Z(t''_{2})g(t+t'-t''_{2}-t_{k'}) \nonumber\\
& \approx & \frac{q^{2}}{\tau} \sum_{m}^{*} \langle \delta o_{n+m} \delta  o_{n} \rangle \int dt'u(t')u(t+t'+m\tau)\nonumber\\
& = & q\langle I_{i} \rangle \sum_{m}^{*} \langle \delta o_{n+m} \delta  o_{n} \rangle U(t+m\tau)
\end{eqnarray}
Here, we have introduced two functions:
\begin{eqnarray}
u(t) & = & \int dt' Z(t')g(t-t')\\
U(t) & = & \int dt' u(t')u(t'+t)
\end{eqnarray}
Their Fourier transforms are
\begin{eqnarray}
\mathscr{F}[u(t)] & = & \tilde{u}(f) \quad = \quad  \tilde{Z}(f)\tilde{g}(f)\\
\mathscr{F}[U(t)] & = & \tilde{U}(f) \quad = \quad  \tilde{u}(f)\tilde{u}^{*}(f)\quad = \quad |\tilde{Z}(f)|^{2}|\tilde{g}(f)|^{2}
\end{eqnarray}
Therefore, the measured shot noise $S_{ID}(f)$ is given by
\begin{eqnarray}
S_{D}(f) & = & 2q\langle I_{i} \rangle \sum_{m\ge 0}^{*} \langle \delta o_{n+m} \delta  o_{n} \rangle cos(m\omega \tau) \tilde{U}(f) \\
& = & 2qF |\tilde{Z}(f)|^{2}\tilde{U}(f) \langle o_{n} \rangle \langle I_{i} \rangle\\
& = & |\tilde{Z}(f)|^{2}S(f)
\end{eqnarray}

\subsubsection{Center frequency $>$ DC}
Suppose that we shift the center frequency to $f_{0}$. The consequence is trivial in the frequency domain. 
\begin{eqnarray}
S_{D}(f) & = & |\tilde{Z}(f-f_{0})|^{2}S(f)
\end{eqnarray}
The effect of the frequency shift in the time domain is also simple.  
\begin{eqnarray}
\mathscr{F}[\tilde{Z}(f-f_{0})] & = & e^{i2\pi f_{0}t}Z(t)\\
Re\Big[\mathscr{F}[\tilde{Z}(f-f_{0})]\Big] & = & cos(2\pi f_{0}t)Z(t)
\end{eqnarray}
Thus, the effect of shifting the center frequency on the detector function $Z(t)$ is merely modulating it with $cos(2\pi f_{0}t)$.

\subsection{Cross Correlation}
\subsubsection{General form of auto correlation}
The shot noise formula given above is as follows.
\begin{eqnarray}
C(t) & = & \langle \delta I(t') \delta I(t'+t) \rangle\\
S(f) & = & 2|\mathscr{F}[C(t)]|  
\end{eqnarray}
This is nothing but the auto correlation function of the current fluctuation. We name this current as current 1, $I_{1}(t) = q_{1}g_{1}(t)*h_{1}(t)$ with $t_{1k}=k\tau_{1}$. Then, we may denote it as follows.
\begin{eqnarray}
C(t)& \Rightarrow & C^{11}(t) \quad = \quad \langle \delta I_{1}(t')\delta I_{1}(t'+t) \rangle \\
{S}(f) & \Rightarrow & S^{11}(f) \quad = \quad 2|\mathscr{F}[C^{11}(t)]|  
\end{eqnarray}
Let us also define the following.
\begin{eqnarray}
G(t)& \Rightarrow & G_{11}(t) \quad = \quad \int dt'g_{1}(t')g_{1}(t+t')\\
F& \Rightarrow & F_{11}(f) \quad = \quad \sum_{m\ge 0}^{*} \frac{\langle \delta o_{1n+m} \delta o_{1n} \rangle}{\langle o_{1n} \rangle}cos(m\omega\tau_{1})
\end{eqnarray}
One can also include the detector as previously discussed.
\begin{eqnarray}
U(t)& \Rightarrow & U_{11}(t) \quad = \quad \int dt'u_{1}(t')u_{1}(t+t')\\
C_{D}^{11}(t) & = & \langle \delta I_{D1}(t')\delta I_{D1}(t'+t) \rangle \\
S_{D}^{11}(f) & = & = \quad 2|\mathscr{F}[C_{D}^{11}(t)]|  
\end{eqnarray}
So far, this is nothing but a change of notation. The benefit from this generalization becomes clear in the followings.

\subsubsection{General form of cross correlation}
We may generalize this formula by considering the cross correlation function between two kinds of currents. Namely, $I_{1}(t) = q_{1}g_{1}(t)*h_{1}(t)$ and $I_{2}(t) = q_{2}g_{2}(t)*h_{2}(t)$, where $h_{1}(t) = \sum_{k=1}^{N_{1}}o_{1k} \delta(t-t_{1k})$ and $h_{2}(t) = \sum_{k'=1}^{N_{2}}o_{2k'} \delta(t-t_{2k'})$. As before, $t_{1k}=k\tau_{1}$ and $t_{2k'}=k'\tau_{2}$. $q_{1} \neq q_{2}$ means two currents consists of two different charged entities. $\tau_{1} \neq \tau_{2}$ means the two currents consists of slots with different duration. However, they should satisfy $N_{1}\tau_{1} = N_{2}\tau_{2}$. To be more concrete, two currents are coming from different contacts which are voltage-biased differently. $o_{1k}$ can obviously be different $o_{2k'}$ in general. The cross correlation of the current fluctuations of these two currents is defined as follow.
\begin{eqnarray}
C_{D}^{12}(t)& = & \langle \delta I_{D1}(t')\delta I_{D2}(t'+t) \rangle
\end{eqnarray}
The detailed calculation follows:
\begin{eqnarray}
C_{D}^{12}(t)
& = & \lim_{\mathscr{T} \rightarrow \infty} \frac{1}{\mathscr{T}}\int_{-\mathscr{T}/2}^{\mathscr{T}/2} dt' \delta I_{D1}(t')\delta I_{D2}(t'+t)\nonumber\\
& = & \lim_{\mathscr{T} \rightarrow \infty} \frac{q_{1}q_{2}}{\mathscr{T}} \sum_{k=1}^{N_{1}}\sum_{k'=1}^{N_{2}} \delta o_{1k}\delta o_{2k'} \int dt'u_{1}(t')u_{2}(t+t'+t_{1k}-t_{2k'})\nonumber\\
& = & \lim_{\mathscr{T} \rightarrow \infty} \frac{q_{1}q_{2}}{\mathscr{T}} \sum_{k=1}^{N_{1}}\sum_{k'=1}^{N_{2}} \delta o_{1k}\delta o_{2k'} U_{12}(t+t_{1k}-t_{2k'})
\end{eqnarray}
, where, obviously, $U_{12}(t)=\int dt'u_{1}(t')u_{2}(t+t')$. With this maximum generality, it is difficult to proceed much further. We have to take some simplifications. However, in general, one is not interested in cross correlating two signals prepared completely differently. Let us say that our main playground is the mesoscopic electric transport, one is oftern interested in cross correlating two currents after partinining a current or two currents with same voltage-biases such as in collision experiments. Hence, without losing much of generality, it can be simplified once we take the case $\tau_{1} = \tau_{2} = \tau$, thus, $N_{1}=N_{2}=N$.
\begin{eqnarray}
C_{D}^{12}(t) & = & \lim_{\mathscr{T} \rightarrow \infty} \frac{q_{1}q_{2}}{\mathscr{T}} \sum_{k,k'=1}^{N} \delta o_{1k}\delta o_{2k'} U_{12}(t+t_{k}-t_{k'})\nonumber\\
& = & \lim_{\mathscr{T} \rightarrow \infty} \frac{q_{1}q_{2}}{\mathscr{T}} \sum_{m=-(N-1)}^{N-1} \sum_{n=[1:N-m], m\ge 0} ^{n=[|m|+1:N], m<0}  \delta o_{1n+m}\delta o_{2n} U_{12}(t+m\tau)\nonumber\\
& \approx & \frac{q_{1}q_{2}}{\tau} \sum_{m}^{*} \langle \delta o_{1n+m}\delta o_{2n} \rangle U_{12}(t+m\tau)\nonumber\\
& = & \frac{q_{1}q_{2}}{\tau}\sqrt{\langle o_{1n} \rangle \langle o_{2n} \rangle}\sum_{m}^{*} \frac{\langle \delta o_{1n+m}\delta o_{2n} \rangle}{\sqrt{\langle o_{1n} \rangle \langle o_{2n} \rangle}} U_{12}(t+m\tau)\nonumber\\
& = & \sqrt{q_{1}q_{2}}\sqrt{\frac{q_{1}\langle o_{1n}\rangle}{\tau}}\sqrt{\frac{q_{2}\langle o_{2n}\rangle}{\tau}}\sum_{m}^{*} \frac{\langle \delta o_{1n+m}\delta o_{2n} \rangle}{\sqrt{\langle o_{1n} \rangle \langle o_{2n} \rangle}} U_{12}(t+m\tau)\nonumber\\
& = & \sqrt{q_{1}q_{2}}\sqrt{\langle I_{1} \rangle\langle I_{2} \rangle}\sum_{m}^{*} \frac{\langle \delta o_{1n+m}\delta o_{2n} \rangle}{\sqrt{\langle o_{1n} \rangle \langle o_{2n} \rangle}} U_{12}(t+m\tau)
\end{eqnarray}
Its power spectrum follows
\begin{eqnarray}
S_{D}^{12}(t) & = & 2\sqrt{q_{1}q_{2}}\sqrt{\langle I_{1} \rangle\langle I_{2} \rangle} \sum_{m\ge 0}^{*} \frac{\langle \delta o_{1n+m} \delta o_{2n} \rangle}{\sqrt{\langle o_{1n} \rangle\langle o_{2n} \rangle}} cos(m\omega \tau) \tilde{U}_{12}(f)\\
& = & 2\sqrt{q_{1}q_{2}}\sqrt{\langle I_{1} \rangle\langle I_{2} \rangle} F_{12}(f)\tilde{U}_{12}(f)
\end{eqnarray}
If $q_{1}=q_{2}=q$, $\langle I_{1} \rangle = \langle I_{2} \rangle$.
\begin{eqnarray}
C_{D}^{12}(t) & = & q\langle I_{i} \rangle \sum_{m}^{*}\langle \delta o_{1n+m}\delta o_{2n} \rangle U_{12}(t+m\tau)\\
S_{D}^{12}(t) & = & 2q\langle I_{i} \rangle \sum_{m\ge 0}^{*}\langle \delta o_{1n+m} \delta o_{2n} \rangle cos(m\omega \tau) \tilde{U}_{12}(f)
\end{eqnarray}

\subsubsection{Cross correlation of electrons from single splitter}
Let us make an exercises on this cross correlation. The simplest case is cross correlation between the transmitted and the reflected currents from a quantum point contact (QPC) in the filling factor 1 of quantum Hall system. A noiseless current $I_{i}(t)=\langle I_{i} \rangle$ is impinged on the QPC. $q_{1}=q_{2}=e$ and wave packets obey the Fermi statistics. The occupation of each slot is, hence,
\begin{eqnarray}
o_{2k} & = & 1-o_{1k} \quad \Rightarrow \quad \delta o_{2k} \quad = \quad -\delta o_{1k} \quad = \quad -\delta o_{k}.
\end{eqnarray}
Therefore the cross correlation of two currents is
\begin{eqnarray}
C_{D}^{12}(t)
& = & -e \sqrt{\langle I_{1} \rangle \langle I_{2} \rangle} \sum_{m}^{*} \frac{\langle \delta o_{n+m}\delta o_{n} \rangle}{\sqrt{\langle o_{1n} \rangle\langle o_{2n} \rangle}} U_{12}(t+m\tau)\\
& = & -e \sqrt{\langle I_{1} \rangle \langle I_{2} \rangle} \sum_{m}^{*} \frac{\langle \delta o_{n+m}\delta o_{n} \rangle}{\sqrt{\langle o_{1n} \rangle\langle o_{2n} \rangle}} U_{11}(t+m\tau)\\
& = & -C_{D}^{11}(t)
\end{eqnarray}
The power spectrum of this cross correlation is 
\begin{eqnarray}
S_{D}^{12}(t) & = & -2e \sqrt{\langle I_{1} \rangle \langle I_{2} \rangle} \sum_{m\ge 0}^{*} \frac{\langle \delta o_{n+m}\delta o_{n} \rangle}{\sqrt{\langle o_{1n} \rangle\langle o_{2n} \rangle}}cos(m\omega \tau) \tilde{U}_{12}(f)\\
& = & -2e F_{12}(f) \tilde{U}_{12}(f) \sqrt{\langle I_{1} \rangle \langle I_{2} \rangle} \\
& = & -2e \sqrt{\langle o_{1n} \rangle\langle o_{2n} \rangle} F_{12}(f) \tilde{U}_{12}(f) \langle I_{i} \rangle \\
& = & -2e \sqrt{T(1-T)} F_{12}(f) \tilde{U}_{12}(f) \langle I_{i} \rangle\\
& = & -2e \sqrt{T(1-T)} F_{11}(f) \tilde{U}_{11}(f) \langle I_{i} \rangle\\
& = & -S_{D}^{11}(t)
\end{eqnarray}
Here, we used $\langle o_{1n} \rangle=T$, $\langle o_{2n} \rangle=1-T$. Hence, making a cross correlation on the transmitted and reflected currents gives the same result as the auto correlation other than the sign. However, in reality, noises of the two detectors are uncorrelated. By averaging the uncorrelated noises, one can get a better signal to noise in the measurement of the shot noise compared to detecting it with one detector given the same averaging time.

\subsubsection{Case of independent tunneling}
For the case of non-correlated tunneling events, $\langle \delta o_{n+m}\delta o_{n} \rangle$ is finite only for $m=0$. Remember $\langle \delta o_{n}^{2}\rangle=T(1-T)$ for independent tunnelings.
\begin{eqnarray}
C^{12}(t) & = & -eT(1-T) \langle I_{i} \rangle U_{11}(t)
\end{eqnarray}
Remember that $ G_{12}(t)=G_{11}(t)=G_{22}(t)$ since the two currents are originating from the same current. Therefore, the power spectrum of the cross correlation reads
\begin{eqnarray}
S^{12}(t) & = & -2eT(1-T) \langle I_{i} \rangle \tilde{U}_{11}(f) \quad = \quad -S^{11}(t)
\end{eqnarray}
, which is nothing but the shot noise of electron currents with negative sign. This negative sign originates from the conservation of current and does not prove anti-correlation among Fermions.

\subsubsection{Effect of time-delay on cross correlation}
In reality, the distances between the tunneling barrier and each detector can differ. This means that we cross correlate signals at different times. Let us name it as time-delayed cross correlation. This time delay can be also intentionally introduced using a proper electronic component. We can easily implement this time delay $t_{d}<0$ as follows. 
\begin{eqnarray}
C_{D}^{12}(t+t_{d})& = & \langle \delta I_{1}(t')\delta I_{2}(t'+t+t_{d}) \rangle
\end{eqnarray}
The correlation function and the power spectrum are easily computed.
\begin{eqnarray}
C_{D}^{12}(t+t_{d}) & = & \sqrt{q_{1}q_{2}}\sqrt{\langle I_{1} \rangle\langle I_{2} \rangle} \sum_{m}^{*} \frac{\langle \delta o_{1n+m} \delta o_{2n} \rangle}{\sqrt{\langle o_{1n} \rangle} \sqrt{\langle o_{2n} \rangle}} U_{12}(t+t_{d}+m\tau)\\
S_{D}^{12}(f, t_{d}) & = & 2\sqrt{q_{1}q_{2}}\sqrt{\langle I_{1} \rangle\langle I_{2} \rangle} \sum_{m\ge 0}^{*} \frac{\langle \delta o_{1n+m} \delta o_{2n} \rangle}{\sqrt{\langle o_{1n} \rangle} \sqrt{\langle o_{2n} \rangle}} cos(\omega (m\tau+t_{d})) \tilde{U}_{12}(f)\nonumber\\
& = & 2\sqrt{q_{1}q_{2}}\sqrt{\langle I_{1} \rangle\langle I_{2} \rangle} F_{12}(f, t_{d})\tilde{U}_{12}(f)\\
& = & \frac{F_{12}(f, t_{d})}{F_{12}(f, 0)}S^{12}(f, 0)
\end{eqnarray}
It is clearly seen that the time delay $t_{d}$ does have an effect on the cross correlation by shifting the phase. To understand it better, let us again take the case of independent tunnelings, namely, keeping only $m=0$ in the sum. 
\begin{eqnarray}
S_{D}^{12}(f, t_{d}) & = & 2\sqrt{q_{1}q_{2}}\sqrt{\langle I_{1} \rangle\langle I_{2} \rangle} \frac{\langle \delta o_{1n} \delta o_{2n} \rangle}{\sqrt{\langle o_{1n} \rangle} \sqrt{\langle o_{2n} \rangle}} cos(\omega t_{d}) \tilde{U}_{12}(f)\nonumber\\
& = & -2eT(1-T) \tilde{U}_{12}(f) cos(\omega t_{d}) \langle I_{i} \rangle\nonumber\\
& = & cos(\omega t_{d}) S_{D}^{12}(f, 0)
\end{eqnarray}
For the center frequency $f=$1MHz, suppose there is a 3-meter difference in the cables for the detectors. Disregarding the dielectric constant of the coaxial cable, $t_{d}=\frac{3m}{3\times 10^{8}m/s}=10$nsec. Then, $cos(\omega t_{d}) \approx  0.998$. (Of course, in reality, the coaxial cable adds extra stray capacitance.) Or, suppose that the two cables are equal in length and that the path-length difference between the tunnel barrier and the detectors is $1$mm. For the Fermi velocity of $10^{5}$ m/s, $t_{d}=\frac{1 mm}{10^{5}m/s}=10$nsec and $cos(\omega t_{d}) \approx  0.998$.

\section*{Acknowledgements}
I acknowledge Nissim Ofek, Moty Heiblum, Oren Tal, Bernd Rosenow, and Izhar Neder for valuable discussions. 

\bibliographystyle{mybibstyle}
\bibliography{reference}

\end{document}